\documentclass[fleqn,twoside]{article}
\usepackage{espcrc2}

\usepackage[dvips]{graphicx}

\usepackage{dcolumn}

\def\sfrac#1#2{{\textstyle{ #1 \over #2}}}

\title{Higher-order perturbation theory for highly-improved actions}

\author{Howard D. Trottier\address{
Simon Fraser University, Department of Physics, Burnaby BC V5A 1S6, Canada}}

\begin{document}

\begin{abstract}
I review techniques and applications of higher-order perturbation theory 
for highly-improved lattice actions.
\end{abstract}

\maketitle

\section{Introduction}
Lattice QCD simulations are routinely done nowadays using highly-improved actions, which are
designed to remove the leading errors arising from the discretization of the continuum theory. 
Improved actions can be designed from both perturbative and nonperturbative considerations. 
In this review I describe techniques for doing the higher-order perturbation theory (PT) calculations 
that are necessary in the design of highly-improved lattice discretizations for gluons, 
staggered quarks, and heavy quarks. Recent results obtained with these methods 
are also reviewed.

Much of the work described here is part of the program of the HPQCD collaboration, 
a major goal of which is to make precision calculations of hadronic matrix 
elements relevant to $b$-physics experiments. In order to fully realize the potential impact 
of these experiments on the parameters of flavor mixing requires calculations of the relevant 
hadronic matrix elements to a few percent accuracy. The CLEO-c program also presents an 
enormous opportunity to validate lattice QCD methods for $b$ physics, by testing 
predictions for analogous quantities in the charm system. 

These stringent requirements for accuracy and timeliness are unlikely to be met without significant 
algorithmic developments, in both the efficiency of unquenched simulations, and in the 
technical challenges posed by lattice PT. The development of an
improved action for staggered quarks \cite{LepageStaggered,Golterman} has at last made 
accurate unquenched simulations feasible \cite{MILC} at dynamical quark masses 
that are small enough to allow for reliable chiral extrapolations to the physical region 
\cite{HPQCD}. This is one particularly striking success of the perturbative analysis of 
lattice discretizations. More generally one must do perturbative matching calculations 
for a wide array of coupling constants, action parameters and hadronic matrix elements.

The scope of the charge to lattice PT is set by two expansion parameters: 
$a\Lambda_{\rm QCD}$, where $a$ is the lattice spacing and $\Lambda_{\rm QCD}$ 
is a typical low-energy scale; and the strong coupling $\alpha_s(q^*)$, evaluated near 
the ultraviolet cutoff ($q^*\sim1/a)$ at which the lattice theories are to be matched onto 
continuum QCD. Affordable unquenched simulations can only be done for lattice spacings 
around 0.1~fm, where these two expansion parameters have about the same value:
\begin{equation}
   \alpha_s(1/a) \approx a\Lambda_{\rm QCD} \approx 0.2 \mbox{--} 0.3 
   \quad  [a \approx 0.1~\mbox{fm}] .
\end{equation}
Hence to reduce systematic errors to  a few percent requires lattice discretizations that are
accurate through ${\cal O}(a^2)$, which is trivial to do by numerical analysis,
and matching of the resulting action and operators must be done through $O(\alpha_s^2)$.

The latter requirement is highly nontrivial, and in fact only a few two-loop PT 
calculations in lattice QCD have ever been done. In many ways perturbative calculations 
with a lattice regulator are much more difficult than with dimensional regularization: 
the Feynman rules for lattice theories are exceedingly complicated, even for the simplest 
discretizations, with a tower of contact interactions that leads to a proliferation of 
Feynman diagrams that are not present in the continuum.  

One has the challenge of doing many two-loop matching calculations, 
and doing these for the most complicated actions yet designed. Moreover one may
anticipate further evolution in the actions that are actually used in simulations, 
as investigations continue into the optimal discretizations, hence one will want to be 
able to redo many perturbative calculations as the actions evolve. 

Fortunately there exists several techniques which make higher-order lattice PT 
more manageable. In this review I survey two very different and somewhat complementary
methods: one for conventional Feynman diagram evaluation of loop corrections, 
and another approach in which one does Monte Carlo simulations of the lattice path integral
in the weak-coupling regime. 

The linchpin of our program for diagrammatic PT is the
automatic generation of the Feynman rules for lattice actions. This method was
developed long ago by L\"uscher and Weisz \cite{LWVertex}, and is described in Sect.~2. 
What is new in our work is that we have aggressively applied this method to higher-order 
calculations for highly-improved actions for gluons, staggered quarks, and heavy quarks.
This is in order to analyze the MILC simulations of (2+1)-dynamical 
quark flavors \cite{MILC}, which were done for the $O(\alpha_s a^2)$-accurate Symanzik 
gluon action, and the tree-level $O(a^2)$-accurate staggered quark action due to 
Lepage \cite{LepageStaggered}, both with tadpole improvement. 

A major effort of the past year has been devoted to a complete third-order determination of 
$\alpha_s(M_Z)$, the first from lattice QCD, which should provide one of the 
most accurate determinations of the strong coupling. This requires a set of two-
and three-loop calculations, which are described in Sects.~3 and 4. 

Some other recent PT calculations are briefly reviewed in Sect.~5; a comprehensive 
review of recent work cannot be done in this short space, so I have mainly restricted 
this summary to work done with automated methods.
Monte Carlo methods for the extraction of perturbative expansions are described 
in Sect. 6. Some conclusions and prospects for the future are found in Sect. 7.

\section{Automatic vertex generation}
Lattice Feynman rules are vastly more complicated than in the continuum, even for 
the simplest discretizations (the continuum four-gluon vertex for instance has six terms, while 
for the Wilson gluon action the expression spans dozens of lines). The complexity of these 
rules grows extremely rapidly even for modest increases in the complexity of the action, 
and with the number of lines in the vertex. A path in the action with 
$\ell$ links will generate a vertex function for $r$ gluon lines with a number of 
terms bounded by \cite{LWVertex}
\begin{equation}
    n_{r,\ell} = 2   \ell (\ell+1) \mbox{\ldots} (\ell + r - 1)  / (r-1)!.
\end{equation}
The ${\cal O}(a^2)$-accurate gluon action has $\ell = 6$, while
for third-order expansions one requires vertex functions with $r=6$ gluons,
which have up to $n_{6,6}=5544$ terms (cancellations lead in practice to 
expressions of about half that size). 

On the other hand remarkably simple algorithms can be developed to automate the generation of the 
Feynman rules, for essentially arbitrarily complicated lattice actions. The method which 
we use combines algorithms of L\"uscher and Weisz \cite{LWVertex} and 
Morningstar \cite{Morningstar}. Other algorithms have been developed by
Panagopoulos and collaborators \cite{Panagop}, and Capitani and Rossi \cite{Capitani}.

The algorithm has a very user-friendly interface. One need only specify the action in an 
abstract form, according to the link paths that it contains. The algorithm then 
Taylor expands the link variables, collects  terms of the desired order in the coupling, 
keeping track of Lorentz and color indices, and  Fourier-transforms the fields. 
A final expression for the vertex function is printed in a 
language that is suitable for use in routines where the Feynman diagrams are 
coded, from a combination of vertex functions, allowing for instance a numerical
integration over the loop momenta. (Automatic generation of the Feynman diagrams
themselves can also be done \cite{QThesis}.)

To turn this prose into a computer program that can handle a very complicated action, 
it is helpful to observe that one can build up the Feynman rules from a convolution 
of the rules for simpler elements \cite{Morningstar}. To this end, let's consider the 
world's simplest ``action,'' consisting of a link in a single direction, summed over lattice sites
\begin{equation}
 {\cal S}^{\rm one\mbox{-}link} \! = \!  
\sum_x U_\ell(x) \! = \! \sum_x
 \exp[i g_0 A_{\hat\ell}(x+ \sfrac12 \hat a_\ell)] .
\end{equation}

We can immediately write down the Feynman rules for this ``action,'' since we need only to
take the Taylor series expansion of the lone exponential. The vertex function $V$ for $r$ gluons is

\begin{eqnarray}
\lefteqn{
 V^{\rm one\mbox{-}link}_{\rm unsymm}
  \left(  \{k_i, \mu_i, a_i\} \right) = - i^r \times
} \nonumber\\ 
\lefteqn{
(2\pi)^4 \delta\left( k_1 + k_2 + \mbox{\ldots} + k_r = k_{\rm tot} \right) \times
}\nonumber\\
\lefteqn{
{g_0^r \over r!} \delta(\hat\mu_1 = \hat\mu_2 = \mbox{\ldots} = \hat\mu_r = \hat\ell)
\times  \left\{ T^{a_1} T^{a_2} \!\! \mbox{\dots} T^{a_r} \right\}
}\nonumber\\
\lefteqn{
\times \exp[i ( \sfrac12 k_1 \cdot a_\ell + \sfrac12 k_2 \cdot a_\ell
+ \mbox{\ldots} + \sfrac12 k_r \cdot a_\ell ) ] ,
}
\label{eq:Vlink}
\end{eqnarray}
where we first compute an unsymmetrized vertex, with the gluon labels assigned 
in a fixed order to the link matrices (the color trace has not yet been taken); a sum over 
permutations of the labels is done once the unsymmetrized amplitude for the complete 
action has been constructed.
The origin of each term in Eq.\ (\ref{eq:Vlink}) is simple:  momentum conservation 
(allowing for momentum transfer $k_{\rm tot}$ into the vertex), a $1/r!$ coming from the
Taylor expansion, and phase factors from the Fourier transforms of the gauge fields. 
The spin $\delta$-function arises because all gluons must be 
polarized along the direction of the single link.

\begin{figure}[htb]
\centerline{\framebox{\includegraphics[scale=0.5]{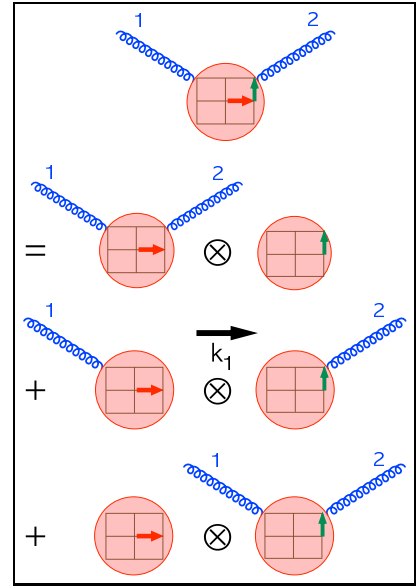}}}
\caption{Generation of an unsymmetrized vertex function by convolution.}
\label{fig:Convolution}
\end{figure}

To build the vertex function for a more complex action
one can convolute the vertices of individual links. This is illustrated 
by another simple example, in Fig.\ \ref{fig:Convolution}, which shows the generation 
of the unsymmetrized vertex function for two gluons ($r=2$) for an ``action'' with two links
($\ell=2$). One simply assigns an ordered set of gluon labels to the ordered set of 
links (so as to respect their non-commutativity), in all possible ways. 

Gluon actions with relatively short link paths do not require a program for convoluting
the rules of individual links; this can easily be done analytically, and L\"uscher and Weisz 
provide an explicit realization of an optimal algorithm in that case.
[They also discuss how to automatically generate Feynman rules for gauge-fixing, although for 
practical purposes this can be done by hand.] In more complex cases however it is 
advantageous to explicitly code the convolution theorem, applied to basic operators 
in the action. Consider for instance the NRQCD action 
\begin{eqnarray}
\lefteqn{
{\cal L}_{\rm NRQCD}  =  \psi^\dagger \left( 1 - \frac{a}{2} \delta H \right)
\left(1 + { \Delta^{(2)} \over 4nM_0 } \right)^n
U^\dagger_4
} \nonumber \\
\lefteqn{
\times \left(1 + { \Delta^{(2)} \over 4nM_0 } \right)^n 
\left( 1 - \frac{a}{2} \delta H \right) \psi
-\psi^\dagger \psi,
}
\end{eqnarray}
where $\delta H$ has many operators
\begin{equation}
   a \delta H  \! = \!\!
   - {c_3  g \over 8M^2 } \mbox{{\boldmath$\sigma$}} \cdot 
    ( {\bf \Delta} \times  {\bf E}  - {\bf E} \times {\bf \Delta} )
   -  { c_4 g \over 2 M } \mbox{{\boldmath$\sigma$}} \cdot {\bf B}
   + \ldots .
\label{eq:deltaH}
\end{equation}
The vertex functions for this extremely complex action can easily be built-up
by convolution of the rules for simple operators, such as $\Delta^{(2)}$ 
and ${\bf E}$ and ${\bf B}$ \cite{Morningstar}. We have done a number of 
calculations for NRQCD using exactly this procedure.

\section{Two-loop renormalized coupling}
The convergence of perturbative  expansions is greatly improved by using a renormalized 
coupling \cite{LepageMackenzie}, such as defined by the static-quark potential, 
according to
\begin{equation}
   V(q) \equiv -4\pi C_F \alpha_V(q) / q^2  .
\end{equation}
We have computed the  connection between $\alpha_V(q)$ and the bare lattice coupling 
$\alpha_0$ for improved actions in two steps \cite{HDTMason}. 
We first follow L\"uscher and Weisz \cite{LWAlpha}, who used a background-field technique 
to compute the relation between $\alpha_0$  and $\alpha_{\overline{\rm MS}}$ for the 
Wilson gluon action (the matching for Wilson and clover fermions was done in 
\cite{Panagop}). We then use the relation between $\alpha_{\overline{\rm MS}}$ 
and  $\alpha_V$, which is known through third-order \cite{Schroder}. 

As in the continuum, background-field quantization reduces the number of independent 
renormalization constants. One computes background- and quantum-field two-point 
functions on the lattice, $\Gamma^B$ and $\Gamma^A$ respectively, where
\begin{eqnarray}
\lefteqn{
   \Gamma^{B}(q,-q)^{ab}_{\mu\mu} = -\delta^{ab} 3 \hat q^2 
   \left[1 - \nu(q) \right] / g_0^2 , 
}\nonumber \\
\lefteqn{
   \Gamma^{A}(q,-q)^{ab}_{\mu\mu} = -\delta^{ab} \hat q^2 
   \left[ 3\left( 1 - \omega(q) \right) + \lambda_0 \right],
}
\end{eqnarray}
where $g_0$ and $\lambda_0$ are the bare lattice coupling and gauge parameter. 
Analogous renormalized quantities $g_{\overline{\rm MS}}$, $\nu_R$, etc.\
are defined in the $\overline{\rm MS}$ scheme. 
The couplings in the two schemes are related by
\begin{equation}
   g_{\overline{\rm MS}}^2 = [1 - \nu_R(q)] / [1 - \nu(q)] \times g_0^2 .
\end{equation}
To solve this equation one must account for the implicit dependence on
the couplings induced by the relation
$\lambda_R = [1 - \omega_R(q)] / [1 - \omega(q)] \times \lambda_0$.

At two-loop order one must evaluate 31 pure-gauge diagrams, and 18 diagrams with
internal fermion lines. In addition one
must compute a number of one-loop diagrams that are induced by the $O(g_0^2)$
expansion of the tadpole renormalization factors in the gluon and quark actions. 

\begin{figure}[t]
\centerline{\includegraphics[scale=0.5]{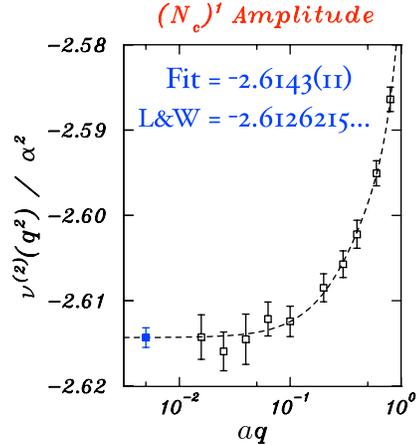}}
\caption{Two-loop results for $\nu(q)$ (Wilson glue).}
\label{fig:FitLW}
\end{figure}

We evaluate loop integrals by numerical integration using VEGAS. This contrasts 
with \cite{LWAlpha,Panagop}, where integrals for unimproved actions were reduced 
analytically to a small set of primitive integrals. In the present case the much more complex
vertices make analytical treatments problematic (see however \cite{Becher}); 
moreover there is no gauge in which the improved gluon propagator is 
diagonal in its Lorentz indices, which further complicates analytical integration. 
We find that numerical integration gives results of sufficient quality in reasonable 
computer time. We do integrations at several values of 
$aq$, and extrapolate to the continuum limit, as illustrated in Fig. \ref{fig:FitLW}.
We require the fit errors in third-order coefficients to be smaller 
than the systematic error from the uncalculated fourth-order 
corrections, for the couplings relevant to simulations.

We also do calculations for several choices of the gauge parameter, to explicitly verify 
gauge-invariance of the final results. This requires the ${\overline{\rm MS}}$ matching 
function $\nu_R$ at two loops for arbitrary gauge parameter. This was done for the 
pure-gauge theory by Ellis \cite{Ellis}, and we have done the two-loop fermionic 
part for arbitrary $\lambda_R$ (this was previously known only in Feynman 
gauge \cite{Panagop}).

The  relation between the couplings is given by
\begin{equation}
   \alpha_0 = \alpha_V(q)
   \left[ 1 - v_1(q) \alpha_V(q) - v_2(q) \alpha_V^2(q) \right] ,
\end{equation}
plus corrections of $O(\alpha_V^4)$, where
\begin{eqnarray}
\lefteqn{
   v_1(q) = (\beta_0/4\pi) \ln(\pi/aq)^2 + v_{1,0} ,
}\nonumber \\
\lefteqn{
   v_2(q) = (\beta_1/ 16\pi^2) \ln(\pi/aq)^2 - v_1^2(q^2) + v_{2,0} .
}
\label{alphaV}
\end{eqnarray}
For the improved actions used by MILC we find
\begin{eqnarray}
\lefteqn{
   v_{1,0} = 3.57123(17) - 1.196(53)\times 10^{-4} \, N_f ,
}\nonumber \\
\lefteqn{
   v_{2,0} = 5.382(39) - 1.0511(51) N_f .
}
\end{eqnarray}

\section{Wilson loops to third order}
One can extract the value of the strong coupling from lattice 
simulations of short-distance observables. To this end
we have evaluated the three-loop PT expansion of Wilson loops \cite{HDTMason}, 
extending a clever approach in \cite{Panagop} in order to reduce the number of
Feynman diagrams. We evaluate a vacuum-to-vacuum transition amplitude, in
the presence of an external ``current'' generated by the Wilson loop itself
\begin{equation}
\langle W_{R,T} \rangle \Bigl\vert_{S_{\rm lat}}
= {\partial \over \partial \rho}
\langle  1  \rangle 
\Bigl\vert_{S_{\rm lat} + \rho W_{R,T}} \quad \left[\mbox{at\ } \rho=0\right],
\end{equation}
using our vertex generators to make the rules for
the ``extended'' action $S_{\rm lat} + \rho W_{R,T}$. This reduces
the number of three-loop diagrams by more than half compared to a ``direct'' 
evaluation of $\langle W_{R,T} \rangle$.

The third-order expansion has 15 gluonic three-loop diagrams, and 19 fermionic ones. 
In addition there are a number of one- and two-loop mean-field counterterm diagrams. 
We quote results for
\begin{equation}
- \! \ln W_{R,T} \! = \! w_0  \alpha_{\rm V}(q^*)
 [ 1 + r_1 \alpha_V + r_2 \alpha_V^2]  \! + \! \ldots.
\label{lnWilsonRen}
\end{equation}
The gluonic parts $r_{i,g}$ are given in Table \ref{table:WilsonLoops}, which 
demonstrates convergence of the renormalized PT expansion through three loops. 

\begin{table}[htb]
\caption{Perturbative expansions for small Wilson loops for the
$O(\alpha_sa^2)$-improved gluon action.}
\begin{tabular}{c|c|c|c|c}
\hline
$R$x$T$    &  $aq^*$   & $w_0$   & $r_{1,g}$   & $r_{2,g}$    \\
\hline
1x1   &   3.33   &   3.0684    & -0.7753(2)     & -0.722(39)   \\    
1x2   &   3.00   &   5.5512    & -0.6202(4)     & -0.407(40)   \\    
1x3   &   2.93   &   7.8765    & -0.5335(8)     & -0.245(44)   \\    
2x2   &   2.58   &   9.1998    & -0.4934(10)   & -0.030(51)   \\ 
\hline   
\end{tabular}
\label{table:WilsonLoops}
\end{table}

\begin{figure}[htb]
\centerline{\includegraphics[scale=0.45]{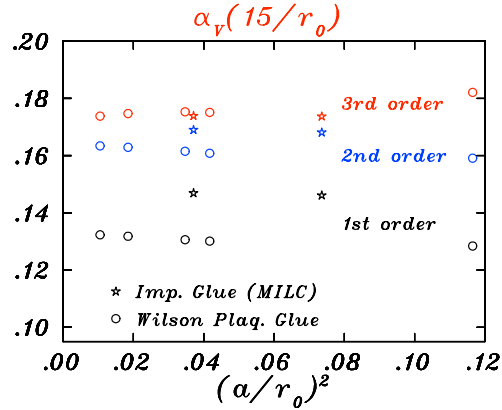}}
\caption{Quenched coupling vs.\ lattice spacing.}
\label{fig:QuenchedCoupling}
\end{figure}

Results for the quenched coupling extracted from simulations at different orders
in the perturbative expansion are shown in Fig.\ \ref{fig:QuenchedCoupling} \cite{Christine}.
This indicates convergence through fourth order, with the results from two gluon actions 
at $n$th-order differing by $O(1)\times\alpha_V^{n+1}$, all the way from $n=1$ through $n=3$.
We are currently analyzing the results for the MILC (2+1)-flavor configurations.

\section{Other recent results}
\begin{figure}[htb]
\centerline{\framebox{\includegraphics[scale=0.525]{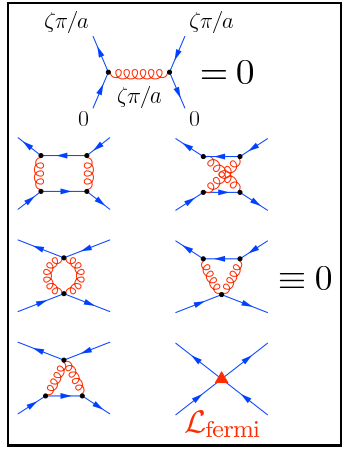}}}
\caption{Taste-changing effects for improved-staggered quarks at tree-level
and one-loop.}
\label{fig:FourFermion}
\end{figure}

The $O(a^2)$-improved staggered-quark action removes taste-changing interactions at 
tree level by suppressing the coupling of quarks to high-momentum 
gluons \cite{LepageStaggered,Golterman}. The ancient problem of a bad 
perturbation theory for staggered quarks has also been shown to be thereby 
eliminated \cite{LeeSharpe}. These effects can be removed at higher orders 
by introducing four-fermion counterterms, cf.\ Fig. \ref{fig:FourFermion}.

Hasenfratz has found that taste-symmetry breaking can be further reduced by additional 
smearing and reunitarization of the gauge links in the staggered-quark action 
\cite{Hasenfratz}. We have found that this improvement is also largely perturbative 
in origin \cite{FollanaMason}. Taste-symmetry violations were measured
from pion splittings in quenched simulations of staggered actions with different link
prescriptions, and calculations of the one-loop four-fermion counterterms were also
done in each case. Perturbation theory correctly accounts for changes
in the pion splittings with changes in the action, 
and the same optimal action is found perturbatively 
and nonperturbatively. Algorithms for doing unquenched simulations with these
actions are also under investigation \cite{HornbostelWoloshyn}.

Another successful PT description of nonperturbative simulation data comes from a 
computation of the one-loop renormalized anisotropy for improved gluon actions \cite{Horgan}. 
Perturbative and nonperturbative determinations of the anisotropy 
were compared over a wide range of lattice spacings and anisotropies, the differences 
in all cases consistent with an $O(1)\times\alpha_V^2$ correction. 

An important application of lattice PT is matching of the Fermilab and NRQCD 
heavy-quark actions and associated currents \cite{Kronfeld,Nobes,Shigemitsu,Gray}.
The first PT matching for the heavy-quark clover interaction was done in \cite{FlynnHill}, 
and new results were presented here by Kayaba \cite{Kayaba}. This is the subject of a 
concerted effort by the HPQCD collaboration; complete one-loop results for the action 
parameters are coming soon \cite{Nobes}. We are also well on the road to the two-loop 
kinetic mass for light and heavy quarks \cite{HDTMason} (see also \cite{PanagopFollana}).

\section{Monte Carlo methods}
There is an attractive alternative to Feynman diagram analysis, which has largely been 
under-exploited. This is to ``do'' PT by doing conventional Monte Carlo evaluations 
of the lattice path integral, but in the ``unconventional'' weak coupling 
phase of the theory \cite{Dimm}. 

\begin{figure}[htb]
\centerline{\includegraphics[width=15pc,height=10pc]{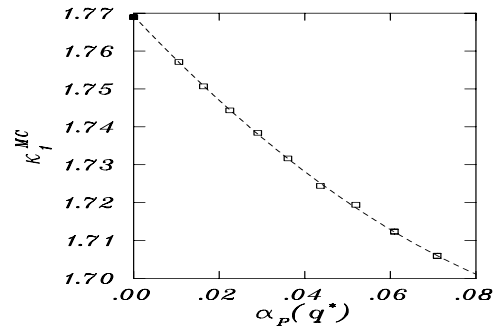}}
\caption{Monte Carlo data for a $5\!\times\!5$ Wilson loop, with 
$\kappa_1 \! \equiv \! -\ln\langle W_{5\times5} \rangle / (20\alpha(q^*))$
[Wilson glue].}
\label{fig:WilsonK1}
\end{figure}

One simulates an observable,
 \begin{equation}
   \langle {\cal O} \rangle 
   = \int [dU_\mu(x)] [d\bar\psi(x) d \psi(x)] \, {\cal O}  \,
   e^{-\beta S_{\rm lat}} ,
\label{eq:PathIntegral}
\end{equation}
over a range of weak couplings, say $\beta\approx9\mbox{--}60$, in a finite (Planck!) box, 
where all lattice momenta are very large (except for possible zero modes, which can be 
eliminated by an appropriate choice of boundary conditions).
One then fits the results to the series
$\langle {\cal O} \rangle  = \sum_n c_n \alpha_V^n(q^*) $.
Third-order expansions of Wilson loops and the static-quark self-energy were obtained in
quenched simulations of  the plaquette action in \cite{QuenchedHighBeta}. 
Some results are shown in Fig.\ \ref{fig:WilsonK1}; 
the intercept of the graph is the leading order coefficient $c_0$, while the slope of 
the curve resolves $c_1$, and its curvature resolves $c_2$. We subsequently did 
the three-loop expansions by Feynman diagrams \cite{HDTMason}, and the 
results are in excellent agreement, see Table \ref{table:HighBeta}. 

\begin{table}[htb]
\caption{Monte Carlo (MC) and PT results for $r_{2,g}$ (Wilson gluon action).}
\label{table:HighBeta}
\begin{center}
\begin{tabular}{c|c|c}
\hline
Loop    &  MC   &  PT    \\
\hline
$1 \times 2$  &  -1.34(8)\ \     &  -1.31(1)\ \      \\    
$1 \times 3$  &  -1.17(9)\ \     &  -1.13(5)\ \      \\    
$1 \times 4$  &  -1.04(9)\ \     &  -1.00(8)\ \      \\    
$1 \times 5$  &  -0.98(11)         &  -1.05(12)           \\    
$2 \times 2$  &  -0.71(8)\ \     &  -0.71(5)\ \       \\    
$2 \times 3$  &  -0.44(9)\ \     &  -0.48(9)\ \      \\    
$3 \times 3$  &  -0.12(9)\ \     &  \  0.15(17)           \\   
\hline 
\end{tabular}
\end{center}
\end{table}

An alternative method has been developed by the Parma group, using an explicit
expansion of the Langevin equations in the bare coupling; they have recently 
obtained the unquenched third-order static-quark self-energy \cite{Parma}. 
We have also recently extracted PT expansions from unquenched weak-coupling
simulations \cite{WongWoloshyn}.

\section{Summary and outlook}
Automatic lattice perturbation theory methods are remarkably simple and powerful.
These have been used to do a number of higher-order calculations for highly-improved actions. 
We have recently done the PT for a third-order determination of 
$\alpha_{\overline{\rm MS}}(M_Z)$ from the MILC simulations of (2+1)-flavors 
of dynamical staggered quarks. Conventional Monte Carlo simulations in the weak-coupling 
regime may offer a simple alternative to diagrammatic PT, which has so far been 
under-exploited. Our primary goal is a two-loop determination of heavy-flavor physics. 
This is an ambitious program but the technology has been proven at the requisite order, 
and one can reasonably expect that much of this work will be done in the next few years.

I am indebted to Peter Lepage, who has provided much of the rationale for
this physics program.  I have done most of my PT calculations with Quentin Mason,
Matthew Nobes, and Peter Lepage. I have benefited from many discussions 
with Christine Davies, Junko Shigemitsu, Paul Mackenzie, Andreas Kronfeld, 
Aida El-Khadra, Richard Woloshyn, and Ron Horgan.



\begin{thebibliography}{99}
\bibitem{Kronfeld} A review of heavy-quark physics is given by A.\ Kronfeld,
these proceedings.
\end{thebibliography}
\end{document}